\magnification=1200
\voffset=-1.5truecm\hsize=16.5truecm    \vsize=24.truecm
\baselineskip=14pt plus0.1pt minus0.1pt \parindent=12pt

\lineskip=4pt\lineskiplimit=0.1pt      \parskip=0.1pt plus1pt 
\def\st{\scriptstyle} 
 

\input eplain


\beginpackages
\usepackage{graphicx}
\endpackages

 
\input blackdvi   

 
\let\a=\alpha   \let\d=\delta \let\e=\varepsilon 
\let\f=\varphi \let\g=\gamma       
      \let\p=\pi  
\let\r=\rho    
   
\let\D=\Delta  \let\G=\Gamma

 
\global\newcount\numsec\global\newcount\numfor 
\gdef\profonditastruttura{\dp\strutbox} 
\def\senondefinito#1{\expandafter\ifx\csname#1\endcsname\relax} 
\def\SIA #1,#2,#3 {\senondefinito{#1#2} 
\expandafter\xdef\csname #1#2\endcsname{#3} \else 
\write16{???? il simbolo #2 e' gia' stato definito !!!!} \fi} 
\def\etichetta(#1){(\veroparagrafo.\veraformula) 
\SIA e,#1,(\veroparagrafo.\veraformula) 
 \global\advance\numfor by 1 
 \write16{ EQ \equ(#1) ha simbolo #1 }} 
\def\etichettaa(#1){(A\veroparagrafo.\veraformula) 
 \SIA e,#1,(A\veroparagrafo.\veraformula) 
 \global\advance\numfor by 1\write16{ EQ \equ(#1) ha simbolo #1 }} 
\def\BOZZA{\def\alato(##1){ 
 {\vtop to \profonditastruttura{\baselineskip 
 \profonditastruttura\vss 
 \rlap{\kern-\hsize\kern-1.2truecm{$\scriptstyle##1$}}}}}} 
\def\alato(#1){} 
\def\veroparagrafo{\number\numsec}\def\veraformula{\number\numfor} 
\def\Eq(#1){\eqno{\etichetta(#1)\alato(#1)}} 
\def\eq(#1){\etichetta(#1)\alato(#1)} 
\def\Eqa(#1){\eqno{\etichettaa(#1)\alato(#1)}} 
\def\eqa(#1){\etichettaa(#1)\alato(#1)} 
\def\equ(#1){\senondefinito{e#1}$\clubsuit$#1\else\csname e#1\endcsname\fi}

 
\def\bb{\hbox{\vrule height0.4pt width0.4pt depth0.pt}}\newdimen\u 
\def\pp #1 #2 {\rlap{\kern#1\u\raise#2\u\bb}} 
 
\def\ins #1 #2 #3 {\rlap{\kern#1\u\raise#2\u\hbox{$#3$}}}

\def\pallina{{\kern-0.4mm\raise-0.02cm\hbox{$\scriptscriptstyle\bullet$}}} 
\def\palla{{\kern-0.6mm\raise-0.04cm\hbox{$\scriptstyle\bullet$}}} 
\def\pallona{{\kern-0.7mm\raise-0.06cm\hbox{$\displaystyle\bullet$}}} 
\def\bull{\vrule height .9ex width .8ex depth -.1ex } 
 
\def\data{\number\day/\ifcase\month\or gennaio \or febbraio \or marzo \or 
aprile \or maggio \or giugno \or luglio \or agosto \or settembre 
\or ottobre \or novembre \or dicembre \fi/\number\year} 
 

\newcount\pgn \pgn=1 
\def\foglio{\number\numsec:\number\pgn 
\global\advance\pgn by 1} 
\def\foglioa{a\number\numsec:\number\pgn 
\global\advance\pgn by 1} 
 
\footline={\rlap{\hbox{\copy200}\ $\st[\number\pageno]$}\hss\tenrm \foglio\hss} 
 
 
\def\sqr#1#2{{\vcenter{\vbox{\hrule height.#2pt 
\hbox{\vrule width.#2pt height#1pt \kern#1pt 
\vrule width.#2pt}\hrule height.#2pt}}}}

 \def\\{\noindent}

\let\dpr=\partial

\def\tende#1{\vtop{\ialign{##\crcr\rightarrowfill\crcr 
              \noalign{\kern-1pt\nointerlineskip} 
              \hskip3.pt${\scriptstyle #1}$\hskip3.pt\crcr}}} 
\def\otto{{\kern-1.truept\leftarrow\kern-5.truept\to\kern-1.truept}} 
 
\font\smfnt=cmr8 scaled\magstep0 
\font\myfonta=msbm10 scaled \magstep0 
\def\math#1{\hbox{\myfonta #1}}

\BOZZA

\vglue.5truecm
{\centerline{\bf ON A FINITE RANGE DECOMPOSITION OF THE RESOLVENT } 
{\centerline{\bf OF A FRACTIONAL POWER OF THE LAPLACIAN}

\vglue0.3cm
{\centerline{\it Revised Version} 

\vglue1cm
{\centerline{\bf P. K. Mitter}
\vglue1cm 
{\centerline{\smfnt Laboratoire Charles Coulomb}}
{\centerline{\smfnt CNRS-Universit\'e Montpellier- UMR5221}} 
{\centerline{\smfnt Place E. Bataillon, Case 070,  
34095 Montpellier Cedex 05 France}} 
\vglue0.5cm
{\centerline{\smfnt e-mail: Pronob.Mitter@umontpellier.fr}}

\vglue1cm
\\{\bf Abstract}: We prove the existence as well as regularity of a
finite range decomposition for the resolvent $G_{\a} (x-y,m^2) =
((-\D)^{\a\over 2} + m^{2})^{-1} (x-y) $,
for $0<\a<2$ and all real $m$, in the lattice ${\math Z}^{d}$ as well
as in the continuum ${\math R}^{d}$ for dimension $d\ge 2$. This resolvent occurs as the
covariance of the Gaussian measure underlying weakly self- avoiding
walks with long range jumps (stable L\'evy walks) as well as continuous spin
ferromagnets with long range interactions in the long wavelength or
field theoretic approximation. The finite range decomposition should
be useful for the rigorous analysis of both critical and off-critical
renormalisation group trajectories. The decomposition for the special
case $m=0$ was known and used earlier in the renormalisation
group analysis of critical trajectories for the above models below the
critical dimension $d_c =2\a$. This revised version makes some
changes, adds new material, 
and also corrects errors in the previous version. It refers to the
author's published article with the same title in J Stat Phys (2016)
163: 1235-1246 as well as to an erratum to be published in J Stat Phys.

\vglue0.5cm
\\{\bf 1. Introduction} 
\numsec=1\numfor=1
\vglue0.5cm
Let $G$ be a positive definite distribution or function on ${\math
R}^{d}$ or ${\math Z}^{d}$. We say that $G$ has a {\it finite range}
decomposition as a sum of functions, called {\it fluctuation
covariances}, 

$$G=\sum \G_j  \Eq(1.1)$$

\\if the following conditions are met:

\vglue0.2cm
\\1. {\it Positive Definiteness}: The functions $\G_j$ are positive definite.
\vglue0.2cm
\\2. {\it Finite Range}: For some integer L with $L>1$ the $\G_j$  have
finite range:

$$\G_j(x)= 0:  \> |x|\ge L^{j} \Eq(1.2)$$ 

\\We will also require an additional property very useful in applications:
\vglue0.2cm

\\3. {\it Regularity} : The functions $\G_j$ are sufficiently
differentiable and satisfy uniform bounds.
This property can be appropriately defined for lattice functions.

\vglue0.3cm
\\There is as yet no general classification of positive definite
functions/distributions for which all three properties listed above
hold.  However the situation is better when we come to the Green's
function or resolvent
of selfadjoint elliptic operators which can be defined by Dirichlet
forms. The simplest example is the resolvent of the laplacian

$$G(x-y, m^2)= (-\D + m^{2})^{-1} (x-y) \Eq(1.3)$$ 

\\Here $\D$ is the usual laplacian in  ${\math R}^{d}$ or the
lattice laplacian $\D_{\math Z ^d}$  in ${\math Z}^{d}$.  It was
proved by Brydges, Guadagni and Mitter in [9] that in this case a finite range decomposition in the
above sense holds on the lattice and the continuum. Moreover various
convergence theorems were proved in [9] with further developments by Brydges and
Mitter in [10]. Brydges and Talarczyk gave in [12]
partial results for finite range decompositions of Green's functions of quite general elliptic operators (including higher
order operators as well as variable coefficients) defined by Dirichlet forms. Properties 1) and 2)
above were proved whereas property 3) (regularity) was proved for only
the simplest elliptic operators with constant coefficients like the laplacian. Adams, Koteck\'y and
M\"uller [4] extended the results in [9] and [12] to discrete second
order elliptic systems with constant coefficients defined by Dirichlet forms
and proved  regularity of their decomposition.  All these papers,
beginning with [9], use an averaging procedure using Poisson kernels
to derive finite range decompositions. On the other hand Bauerschmidt [5] 
has given a different and novel theory which exploits the finite
propagation speed for hyperbolic systems in order to obtain  finite range decompositions, including
regularity estimates,  for  Green's
functions of elliptic operators, including elliptic systems and
variable coefficients, defined by Dirichlet forms.

\vglue0.3cm

Let now $\a$ be
a real number such that $0<\a<2$. Define the resolvent

$$G_{\a} (x-y,m^2) = ((-\D)^{\a\over 2} + m^{2})^{-1} (x-y) \Eq(1.4)$$ 

\\This is, amongst other things, the resolvent of a (stable) L\'evy
walk, $\a$ being the L\'evy-Khintchine parameter with $m^{2}$ being the
inverse of the killing time of the walk. But it also appears
in other contexts which we will explain later.
When  $m=0$, the Green's function $G_{\a} (x-y;0)$ has the
convergent integral representation for $0<\a<2$ 

$$G_{\a} (x-y,0)= {\sin \p {\a\over 2} \over \p}  \int_{0}^{\infty} ds\>
 s^{-{\a\over2}} G(x-y,s)    \Eq(1.5) $$

\\where on the right hand side $G(\cdot, s)$ is the resolvent of the
laplacian as in \equ(1.3). This can be verified by Fourier transforms
and change of variables (see Lemma 2.2 below). 

\vglue0.3cm

\\{\it Remark}:   Notice that $G_{\a}(x-y,0)$
is well defined both in the lattice and in the continuum for $d\ge 2$
 provided $0<\a<2$. It is also well defined for $d=1$ if we restrict
$\a$ to the range $0<\a<1$. With these restrictions $G_{\a}(x-y,0)$
is a distribution in the continuum (a Riesz potential ) as follows from the following expression:

$$G_{\a, c}(x-y,0)= c(\a, d) |x-y|^{-(d-\a)}  \Eq(1.51) $$

\\The subscript $c$ on the left refers to a continuum expression and the
constant $c(\a, d)$ depends on $\a$ and $d$.

By substituting in \equ(1.5)  the known finite range
decomposition for $G(x-y,s)$ we obtain as in [9] the finite range
decomposition for $G_{\a} (x-y,0)$ with the requisite regularity
properties. The question is what happens if $m\neq 0$. Is there a
finite range decomposition with requisite regularity properties for 
$G_{\a} (x-y,m^2)$ for $m\neq 0$? In this paper we will show that the
answer is in the affirmative. There is a spectral weight  $\r_{\a} (s,
m^{2})$ which collapses to the known one for $m^{2}=0$ and whose
properties are discussed later such that

$$ G_{\a}(x-y,m^{2})= \int_{0}^{\infty} ds\>   \r_{\a} (s, m^{2}) G(x-y,s)    \Eq(1.52) $$

\\Substitution of the known finite range decomposition  for $G(x-y,s)$
in \equ(1.52) will then lead to the results of this paper. 

\vglue0.3cm
The following Theorem holds both in the continuum ${\math R}^{d}$ as
well as in the lattice ${\math Z}^{d}$. However it will be most useful
when used in lattice Renormalisation Group (RG) analysis. Therefore
for definiteness we will work with the lattice Laplacian 
$\D=\D_{{\math Z}^{d}} $ .

\vglue0.2cm

\\The theorem will be first stated in a form using rescaled fluctuation
covariances, the unrescaled version will then appear as a
corollary. Both versions are useful.

\vglue0.2cm

\\{\bf Theorem 1.1}: 
\vglue0.2cm

\\Let $L=3^{p}$, $p\ge 2$. Let $\e_{j}=L^{-j}$, $j\ge 0$. 
Let $G_{\a}(x-y, m^{2})$ be
defined as in \equ(1.4).  Let $0 < \a <2 $ and $d\ge 2$.
 Let

$$[\f]= {d-\a \over 2} \Eq(1.6)$$ 

\\Let $m$ be any real number.  Then there exist positive definite functions 

$$\G_{j,\a} (\cdot, m^{2}) : (\e_j {\math Z}) ^{d}\rightarrow {\math R} \Eq(1.7)$$

\\of finite range

$$\G_{j,\a} (x, m^{2}) =0 : \>  |x|\ge L \Eq(1.8)$$

\\such that for all $x,y \in {\math Z}^d$

$$G_{\a} (x-y, m^{2}) =\sum_{j\ge 0} \> L^{-2j[\f]} \> \G_{j,\a} ({x-y\over L^{j}}, L^{j\a}m^{2}) \Eq(1.9) $$ 

\\We have the regularity  bounds, for all $j\ge 2$ and $0\le q\le j$, and all $p\ge 0$,

$$|| \dpr _{\e_j} ^{p} \G_{j,\a} (\cdot, m^{2}) ||_{L^{\infty}
((\e_q{\math Z})^{d})} \le c_{L, p,\a} (1+ m^{2})^{-2} \Eq(1.10) $$

\\For $j=0,1$ and $0\le q\le j$ we have the bound 

$$|| \dpr _{\e_j} ^{p} \G_{j,\a} (\cdot, m^{2}) ||_{L^{\infty}
((\e_q{\math Z})^{d})} \le c_{L, p,\a} (1+ m^{2})^{-1} \Eq(1.101) $$

\\In the above $\dpr _{\e_j} = \dpr _{\e_j, e_k} ,\> k=1,..,d$ is a forward lattice partial derivative with
increment $\e_j$  and in any particular direction $e_k$ in the lattice
$(\e_j {\math Z}) ^{d} $.  Moreover
$\dpr_{\e_j} ^{p} $ is a multi-derivative of order $p$ defined as in the
continuum  but now with lattice forward derivatives. $e_1,....,e_d$
are unit vectors which give the orientation of ${\math R}^{d}$ as well
as the orientation of all embedded lattices $(\e_j {\math Z}) ^{d}
\subset {\math R}^{d}$. By construction the lattices are nested in an
obvious way. The constant $c_{L, p,\a}$ depends on $L, p, \a$. It depends
implicitly on the dimension $d$. 

\vglue0.3cm

\\Moreover there exist
$C^{\infty}$ positive definite continuum functions $\G_{c_{*},\a} (\cdot, m^{2})$  in ${\math R}^{d}$,
of finite range $L$,  such that as $j\rightarrow \infty$ we have

$$\dpr _{\e_j} ^{p} \G_{j,\a} (\cdot, m^{2}) \rightarrow \dpr_{c}^{p}\G_{c_{*},\a} (\cdot, m^{2})   \Eq(1.11) $$ 

\\in $L^{\infty} ((\e_q{\math Z})^{d})$, for all $p\ge 0$ and $ q\ge 0$. The
convergence is exponentially fast so that for $j\ge 2$ and sufficiently large

$$||\dpr _{\e_j} ^{p} \G_{j,\a} (\cdot, m^{2})
-\dpr_{c}^{p}\G_{c_{*},\a} (\cdot, m^{2})||_{L^{\infty}((\e_q{\math
 Z})^{d})}  \le  c_{L, p,\a} (1+ m^{2})^{-2}L^{-{j\over2}}    \Eq(1.112) $$ 

\\In the above $\dpr_{c}^{p}$ is
the continuum multiple partial derivative in ${\math R}^{d}$ in the same
directions as in the multiple lattice partial derivative $\dpr _{\e_j} ^{p}$. The
above statement has a transcription using Fourier transforms as shown
in the convergence proof given in [BM]. 

\vglue0.3cm

The finite range decomposition \equ(1.9) and the regularity bounds
\equ(1.10) of Theorem 1.1 have the following immediate corollary using unrescaled
fluctuation covariances:

\vglue0.3cm

\\{\bf Corollary 1.2} :

\vglue0.3cm

\\There is a finite range decomposition for $x,y\in {\math Z}^{d}$

$$G_{\a} (x-y, m^{2}) =\sum_{j\ge 0} \> \tilde \G_{j,\a} (x-y, m^{2})  \Eq(1.12) $$ 

\\where the positive definite functions $\tilde\G_{j,\a} (x, m^{2}) $ 
of finite range $L^{j+1}$ are defined by 

$$\tilde\G_{j,\a} (x, m^{2})= L^{-2j[\f]} \> \G_{j,\a} ({x\over L^{j}}, L^{j\a}m^{2}) \Eq(1.13)$$ 
 
\\and satisfy the regularity bounds:

\\for $j\ge 2$,

$$ ||\dpr_{{\math Z} ^{d}}^{p} \tilde \G_{j,\a} (\cdot,
m^{2})||_{L^{\infty} ({\math Z}^{d})}  \le
c_{L, p,\a} (1+ L^{j\a} m^{2})^{-2} L^{-(2j[\f] +pj)}    \Eq(1.14)$$ 

\\and for $j=0,1$ 

 $$ ||\dpr_{{\math Z} ^{d}}^{p} \tilde \G_{j,\a} (\cdot,
m^{2})||_{L^{\infty} ({\math Z}^{d})}  \le
c_{L, p,\a} (1+ L^{j\a} m^{2})^{-1} L^{-(2j[\f] +pj)}    \Eq(1.141)$$ 

\vglue0.3cm
\\{\bf Coarse graining}: 

\vglue0.3cm

\\The constants $c_{L, p, \a} $ depend on the
scale $L$. Such  a dependence occurs because in the main in
intermediate steps of the proof (Section3) we have used results in [9]
where such a dependence occurs.
In order to get scale independence in the
constants we can pass, following Brydges [29] and Bauerschmidt [30],  to a coarser
scale $L'$ and redefine fluctuation covariances by summing over the
intermediate scales.  Let $r$ be a positive  integer and let $L'=L^{r}$
be the coarse scale. For $L$ fixed we can make $L'$ large by making
$r$ large. Now define for $j\ge 0$ the coarse scale fluctuation covariances as
follows:

$$\tilde \G'_{j,\a} (\cdot,m^{2})= \sum_{l=0}^{r-1} \tilde \G_{l+jr,\a} (\cdot,
m^{2})  \Eq(1.1411)$$ 

\\We now get the coarse scale finite range decomposition

$$ G_{\a} (\cdot , m^{2}) = \sum_{j\ge 0} \tilde\G'_{j,\a} (\cdot,m^{2})  \Eq(1.1412) $$

\\with

$$\tilde \G'_{j,\a} (x-y,m^{2}) =0,\> |x-y|\ge (L')^{j+1} \Eq(1.142)$$

\\Now  it is easy to prove that for a fixed  $L$   the coarse grained
fluctuation covariances satisfy the same bounds as above with new
constants $c'_{L,p,\a}$ that are independent of the
coarser scale $L'$ for $0 <\a <2$ and all $d\ge 2$.  For $d=2$ there
is no $\log L' $ as the
dimension $[\f]={d-\a\over 2}$ remains positive for $d=2$ because $0 <
\a < 2$.  For completeness we give the proof in Appendix A.

\vglue0.3cm

\\{\it Remark}: Coarse graining of estimates in [9]  leads to
independence of constants with respect to the coarse scale $L'$ for
$d\ge 3$, as first shown by Bauerschmidt in [30]. For $d=2$ an
additional $\log L'$ dependence was found [30].

\vglue0.3cm
\\The proof of this theorem is given in the following sections. 
\vglue0.3cm

There have been many applications  in recent years of finite range
decompositions, mostly in the
context  of the mathematical analysis of  Wilson's  Renormalization
Group [21] used to study non-linear 
perturbations of Gaussian measures and their scaling
and continuum limits (see some of the references cited below).
Since the resolvent as well as the
summands in the finite range decomposition are positive definite they qualify as covariances
of Gaussian measures. Correspondingly we have a decomposition of a
Gaussian random field as a sum of independent gaussian random fields known as
fluctuation fields. Because 
of the finite range property the fluctuation fields become
uncorrelated beyond a certain finite distance. This enables us to get
rid of the machinery of cluster expansions in the control of the
fluctuation integration which is an essential step in  renormalization
group (RG) analysis and
brings us closer to hierarchical models. 

\vglue 0.3cm
In the RG approach to scaling and continuum limits the control of the critical RG trajectory is
paramount. Although the finite range expansion of the resolvent is
very useful, only the case $m=0$ is strictly
necessary. In fact this suffices for the analysis leading to the existence 
of the critical parameters and the proof of existence of the stable manifold.
The asymptotic properties of critical correlation functions lead to corresponding critical
exponents. However in order to control critical exponents like that
for the susceptibility, correlation length or specific heat  for ferromagnetic systems or self-avoiding
walks we need to consider also off-critical trajectories. This is
a small mass perturbation of the critical trajectory. In the Landau-Ginzburg-Wilson picture  the
bare mass squared  is the temperature and we are approaching the critical
temperature which is the critical mass squared. The bare mass squared 
can be written as the sum of two pieces. The first piece will be taken
to be the resolvent parameter. The resolvent is the covariance of the
underlying Gaussian measure. This will play the role of the renormalised mass
squared which can be defined to be the susceptibility (renormalisation
condition at zero momentum ). The other piece is kept in the interaction and is
a small perturbation. At the
critical point the renormalised mass vanishes. This corresponds to the
bare mass approaching its critical value. This is at the basis of the
calculation of the critical exponent $\g$ for the susceptibility. This was explained
clearly by K. G. Wilson in [22] in his paper on the Feynman-graph
calculation of critical exponents in the $\e$ expansion. Modern rigorous RG applications of
this scheme for  short range ferromagnets in the critical dimension
can be found in some of the references cited below.

\vglue0.3cm
\\In order to
carry out the RG analysis we would need a finite range decomposition
of resolvents when the resolvent mass parameter  $m\neq 0$. In the case of
self-avoiding simple walks (SAWs) in the critical dimension $d=4$ or the
classical continuous $n$-component spin system with short range interaction in the
critical dimension $d=4$ (the so called n-component  $\f^{4}_4$ model),  the known finite
range decompositions [9, 5] of the resolvent \equ(1.3) play an important role
in the rigorous renormalisation group analysis successfully accomplished by
Bauerschmidt, Brydges and Slade in [6, 7, 8] and multiple references therein. In
particular the logarithmic corrections to mean field critical
behaviour, known in the theoretical physics literature for a very long time,
were successfully obtained for the  susceptibility.  Another application of the
finite range decomposition of [9]  is in the late Pierluigi
Falco's important papers on the Kosterlitz-Thouless phase transition
[14, 15].  Furthermore, Dimock [13] used finite range decompositions in
his RG proof of the infinite volume limit for the dipole gas.
 Of course the use of finite range decompositions is not indispensable. For example, the same logarithmic 
corrections were rigorously derived earlier in the RG framework of Gawedzki
and Kupiainen [16] by Hara [17] and Hara and Tasaki [18] in the case of
infrared $\f^{4}_4$ with scalar ($n=1$)  $\f$.  

\vglue0.3cm
However there is another class of problems where the underlying (or
unperturbed) Gaussian
process has as covariance the resolvent \equ(1.4).  Examples are
weakly self avoiding walks with long range jumps with the jump
distribution given by the L\'evy- Khintchine  formula or continuous
spin ferromagnets with long range interaction in the long wave length approximation. For $0<\a<2$
and dimension $d<4$ the upper critical dimension in these cases is
$d_c=2\a$, as shown by Aizenman and Fernandez [2]. Thus for $d<d_c$ , $\e=2\a-d$ can be taken to be a small
parameter. This is at the basis of Fisher, Ma and Nickel's $\e$
expansion computations in [23] of critical exponents for long range
ferromagnets in the long wave length approximation. The critical
RG trajectories below the critical dimension have been controlled 
in the field theoretic version of long range ferromagnets by Brydges,
Mitter and Scoppola in [11]
in the continuum , and for the self avoiding
case in the lattice by Mitter and Scoppola in [ 20] leading in both
cases to proof of existence of non-trivial RG fixed
points as fist conjectured by Fisher et al in [23].
Abdesselam proved for the model studied in [11] the existence of a RG trajectory joining the
unstable Gaussian fixed point to the nontrivial attractive fixed point.
An introductory  review of
RG analysis of critical long range ferromagnets below the critical
dimension  and their continuum limits can be found in [19]. 
The elementary  spin fields have classical critical exponents
dictated by the unperturbed Gaussian measure, see [19], which confirms the
conjecture in [23].
The existence of an anomalous  critical exponent of a
composite spin field is proved in a  hierarchical version of
[11] in [3]. In [20] only the $m=0$ case of Theorem 1.1 was exploited.
However, as mentioned earlier,  in order to control the critical 
exponents such as those for the susceptibility, correlation length
(defined appropriately for long range systems) or the specific heat we need to consider off-critical RG trajectories. Whence the
need for the finite range decomposition including the case $m\neq 0$
given by Theorem 1.1 and its Corollary 1.2.

\vglue0.3cm
\\{\it Remark}: 

\vglue0.3cm

\\ The above statement of Theorem 1.1  corrects errors  in the
published version [26]. It incorporates the corrections given in
[27]. For $j\ge 2$ the bounds remain unchanged except that we have
added an $L$ dependence explicitly to the constants. For $j=0, 1$ the
$(1+m^2)^{-2}$ decrease in \equ(1.10) is replaced by $(1+m^2)^{-1}$ in
\equ(1.101). The Corollary 1.2 is likewise affected.
How to get rid of the scale dependence in
constants by coarse graining was explained in [27]. Some of these errors were
pointed out by G. Slade in [28].
In [28]  a version of the finite
range decomposition given in Theorem 1.1 has now been used in the study of
critical exponents proposed above. The bounds on fluctuation
covariances are similar to those in Theorem 1.1 but not quite the
same. In particular in [28] a supplementary $(1+m^2)^{-1}$ term occurs in  bounds
 for all fluctuation covariances and not just for the cases $j=0,1$ as
 in Theorem 1.1.

\vglue0.5cm
\\{\bf 2.  An Integral Representation} 
\numsec=2\numfor=1

The proof of Theorem $1.1$, which will be given in the following
section,  will make use of an integral representation for the
resolvent \equ(1.4) valid for all real values of the resolvent
parameter $m$. 

\vglue0.3cm

\\Define

$$ f_{\a} (t, m^{2}) =(t^{\a\over 2} +m^{2})^{-1} \Eq(2.2) $$ 

\\where $0<\a<2$, $m$ is any real parameter and $t>0$.

\vglue0.3cm
\\{\bf Proposition 2.1}
\vglue0.3cm

The function $ f_{\a} (t, m^{2})$ , $0<\a<2$, with the restriction
 $t>0$,  satisfies the following integral representation:

$$ f_{\a} (t, m^{2})=\int_{0}^{\infty} ds\> {1\over s+t} \>\r_{\a} (s, m^{2}) \Eq(2.3)$$

\\where

$$\r_{\a} (s, m^{2}) = {\sin\p\a/2\over \p} \> {s^{\a/2}\over s^{\a}
+m^{4} + 2 m^{2} s^{\a/2} \cos\p\a/2} \Eq(2.4)    $$

\\We have the bound

$$0\le \r_{\a} (s, m^{2})\le c_{\a}\> {s^{\a/2} \over s^{\a} + m^{4}} 
\Eq(2.5) $$

\\where the constant

$$c_{\a}= {\sin\p\a/2\over \p}\> {1\over 1-|\cos\p\a/2|}  >0 \Eq(2.6)$$

\\is finite for all $\a$ such that $0<\a<2$. The integral in \equ(2.3)
converges uniformly in $m$. For $m=0$ and $t>0$  we recover the representation 

$$ f_{\a} (t, 0) = {\sin\p\a/2\over \p}\> \int _{0}^{\infty} ds \>
s^{-\a/2} {1 \over s+t}  \Eq(2.7)$$

\vglue0.3cm

\\{\it Remark 1}: Formula \equ(2.3) which was  discovered independently
by the author was actually given earlier by K. Yosida in
[25]. Yosida attributes the formula to T. Kato.

\vglue0.3cm

\\{\it Remark 2}: The folllowing proof is a shortened version of the
previous one and was given in [26].

\vglue0.3cm

\\{\bf Proof of Proposition 2.1}:

\vglue0.3cm

\\Let  $t>0$. Let ${\cal C}'$ be a closed contour taken clockwise enclosing $-t$ in the complex cut $s$-plane ${\math C}/[0, \infty]$.
Let $(-s)^{\a\over 2}$ be the branch given by $(-s)^{\a\over 2} =|s|
e^{i\pi{\a\over 2}\theta}$ with $-\p<\theta<\p$. Since $0<\a<2$,
$(-s)^{\a\over 2} +m^{2}$ cannot vanish inside ${\cal C}'$. Therefore
by Cauchy's residue theorem 

$$f_{\a} (t, m^{2})= -{1\over 2{\p} i} \int_{{\cal C}'} ds \> {1
\over s+t } \>{1\over (-s)^{\a/2 } +m^{2}  } \Eq(2.141)$$

\\We define the contour $\cal C_{\r}$, as follows:  Let
$\r <\infty$ be real. The contour starts at $+\r$, goes counter clockwise parallel to
the real axis with $\arg(-s)= -\p$ in the upper half $s$-plane, circles the origin counter
clockwise, and then goes out to $+\r$ parallel to the real axis
with $\arg(-s)=\p$  in the lower half plane.  On the circle $-s={\d}
e^{i\theta} , -\p\le \theta\le \p$ with $\d<t$.  At the end we take
the limit $\r \rightarrow \infty$ so that 
${\cal C}_{\r} \rightarrow {\cal C}$, the well known Hankel contour, see e.g. [24]. 
This is illustrated below where $\r\rightarrow \infty$.

\vglue0.3cm

\includegraphics[height=5cm,width=8cm,bb=1 1 517 288]{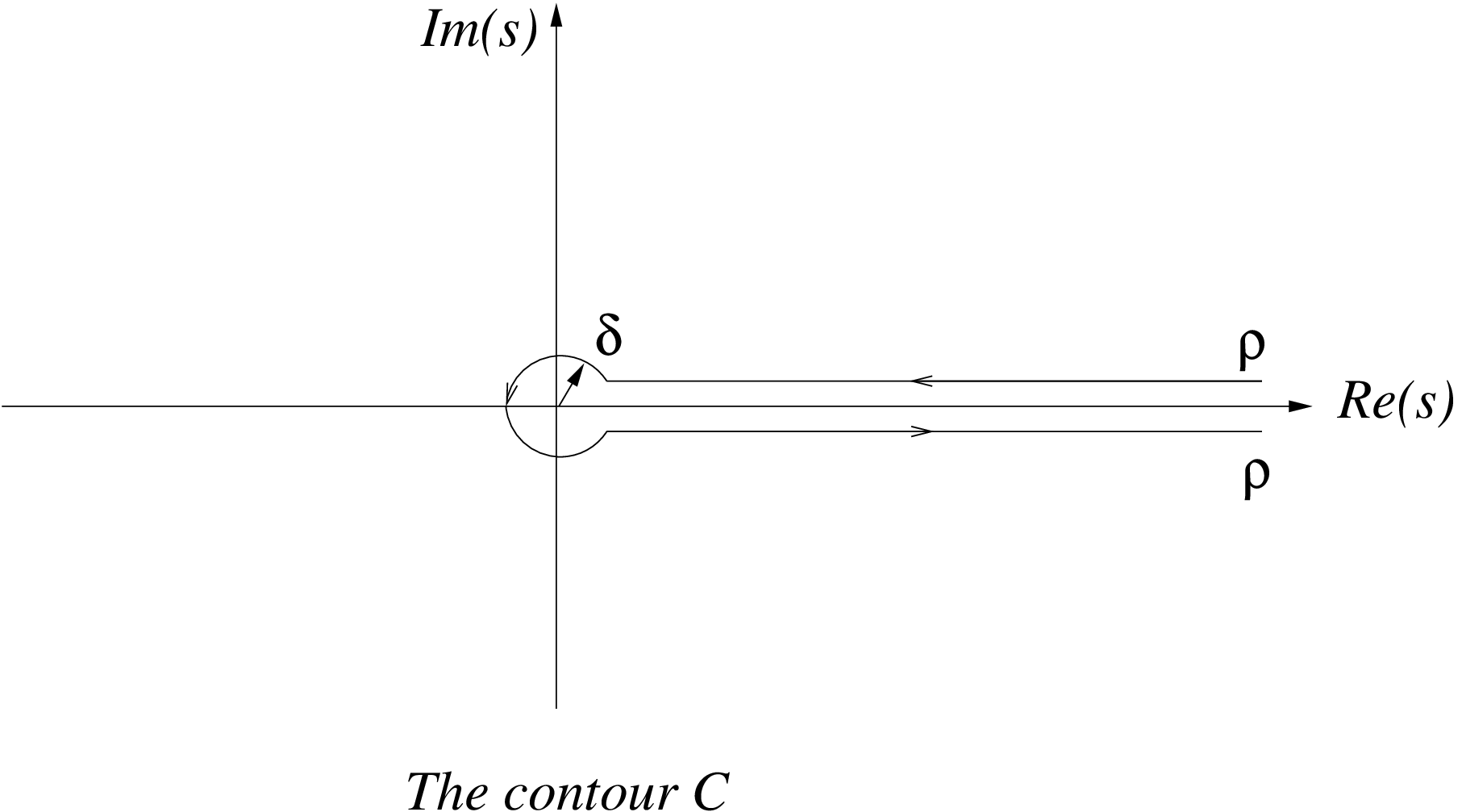}
\vglue0.3cm

\\We now deform ${\cal C}'$ so that it consists of the arc  of a circle of
radius $R$ centered at $-t$ and taken clockwise, and whose two
extremities in the upper and lower half plane then join the 
contour ${\cal C_{\r}}$ with finite $\r$ which
goes around the cut. This is illustrated in the figure below.

\vglue0.5cm

\includegraphics[height=5cm,width=6.5cm,bb=0 0 755 582]{Mitter_contour.pdf}
\vglue0.3cm

\\We now let $R \rightarrow \infty$. This entails that $\r\rightarrow \infty$.
The contribution from the
circular part vanishes and we are left with

$$f_{\a} (t, m^{2})= -{1\over 2{\p} i} \int_{{\cal C}} ds \> {1
\over s+t } \>{1\over (-s)^{\a/2 } +m^{2}  } \Eq(2.142)$$

\\We will prove the proposition by  evaluating the contour integral in
\equ(2.142) as follows:

$$f_{\a} (t, m^{2}) =I_{1,\d} +I_{2,\d}
+ I_{3,\d} $$

\\where

$$I_{1,\d}=- {1\over 2{\p} i}   \int_{+\infty} ^{\d} ds\> 
  {1\over s+t } \>{1\over  e^{-i\p \a/2} s^{\a/2 } +m^{2}  }    $$

$$I_{2,\d} =- {1\over 2{\p} i} \int_{\d} ^{\infty} ds\> 
{1\over s+t } \>{1\over  e^{i\p \a/2} s^{\a/2 } +m^{2}  }   $$

$$I_{3,\d} =\int_{-\p} ^{\p} d\theta \>i\d\> e^{i\theta}  {1 \over t+
\d e^{i\theta} } \>{1\over  (\d e^{i\theta}) ^{\a/2 } +m^{2}  } 
$$

\\We have $I_{3,\d} \rightarrow 0$ as $\d\rightarrow 0$. This is true
for $m^{2}> 0$ and also for $m^{2}=0$ because $0<\a<2$. Letting
$\d\rightarrow 0$  in the sum $I_{1,\d} + I_{2,\d}\rightarrow 0$
we get

$$f_{\a} (t, m^{2})= {1\over 2{\p} i} \int_{0}^{\infty}  ds \> 
{1\over s+t } \>\Bigl( {1\over   e^{-i\p\a/2}s^{\a/2 } +m^{2}  }   
-{1\over   e^{i\p\a/2}s^{\a/2 } +m^{2} }  \Bigr) $$

$$= {1\over 2{\p} i} \int_{0}^{\infty}  ds \> 
{1\over s+t } \> {(e^{i\p\a/2}-e^{-i\p\a/2} ) s^{\a/2 }\over s^{\a}
+m^{4} +2 m^{2} s^{\a/2} \cos \p\a/2 } $$

\\whence we obtain the integral representation of Proposition 2.1

$$ f_{\a} (t, m^{2})=\int_{0}^{\infty} ds\> {1\over s+t} \>\r_{\a} (s, m^{2}) $$

\\where

$$\r_{\a} (s, m^{2}) = {\sin\p\a/2\over \p} \> {s^{\a/2}\over s^{\a}
+m^{4} + 2 m^{2}\> s^{\a/2} \cos\p\a/2} $$

\\Clearly $\r_{\a} (s, m^{2})\ge 0$. We will now prove the upper bound
stated in the Proposition. Let

$$d_{\a}(s, m^{2}) = s^{\a}+m^{4} + 2 m^{2}\> s^{\a/2} \cos\p\a/2 $$

\\ be the denominator in the above formula for the spectral weight
$\r_{\a} $. We have

$$\eqalign{d_{\a}(s, m^{2}) &\ge  s^{\a}+m^{4} - 2 m^{2}\> s^{\a/2} |\cos\p\a/2|\cr
&\ge  s^{\a}+m^{4} -  (m^{4} + s^{\a}) |\cos\p\a/2|\cr
&\ge c'_{\a}(m^{4} + s^{\a})\cr} $$ 

\\where

$$c'_{\a}= 1- |\cos\p\a/2| >0$$

\\since $0<\a< 2$. Hence we obtain the bound for the spectral weight

$$0\le \r_{\a} (s, m^{2})\le c_{\a} {s^{\a/2} \over  m^{4} + s^{\a}}$$

\\where

$$c_{\a}={\sin\p\a/2\over \p} {1\over 1- |\cos\p\a/2|} >0 $$

\\The proof of Proposition 2.1 is complete. \bull

\vglue0.5cm

\\{\bf 3. Proof of Theorem 1.1}
\numsec=3\numfor=1

\vglue0.3cm

\\The proof reposes on Proposition 2.1 and results already obtained in
[9, 10, 5, 27]. This proof corrects that in [26] and incorporates the content
of [27].  We summarise first  only what we need from [9] and
[10] and [27].  
It was proved in [9] that the resolvent of the Laplacian in ${\math
Z}^{d}$

$$G(x-y, s)= (-\D + s)^{-1} (x-y) \Eq(3.1)$$ 

\\with $d\ge 3$ and  $ s\ge 0$    satisfies the finite range expansion

$$G(x-y,s)= \sum_{j\ge 0} \>  L^{-j(d-2)} \> \G_{j} ({x-y\over L^{j}},\>
L^{2j} s ) \Eq(3.2)$$ 

\\where the summands $\G_j$ 

$$\G_j:\> {(\e_j\math Z})^{d} \rightarrow {\math R}  \Eq(3.3)$$ 

\\are positive definite, of finite range $L$  and satisfy  

\\for all integers $j\ge 2$ and $0\le q\le j$ 
and all integers $p\ge 0$ the bounds

$$|| \dpr _{\e_j} ^{p} \G_{j} (\cdot, s) ||_{L^{\infty}
((\e_q{\math Z})^{d})} \le   c_{L, p}   (1+s) ^{-2}    \Eq(3.4) $$

\\where $\e_{j}= L^{-j}$. For $j=q=0$, the above bound is replaced by

$$|| \dpr _{\e_0} ^{p} \G_{0} (\cdot, s) ||_{L^{\infty}
({\math Z})^{d} } \le c_{L, p}{1\over 1+s}    \Eq(3.40) $$

\\The lattice derivatives are as defined after \equ(1.101). In
particular $ \dpr _{\e_0}=\dpr_{{\math Z}^{d}} $. The
constant $c$ in the exponent in \equ(3.4) is of $O(1)$ and independent of $L, j, q, p$. The
constant $c_{p} $ depends on $p$  but is independent $j, q$. It can be verified  that the constant
$c_p$  is actually independent of $L$. They depend
on the dimension $d$.
The bound \equ(3.40)  is proved in  Theorem 5.5 of [9] together with lattice
Sobolev embedding. The bound \equ(3.4) is a slight extension of this
bound. For completeness we give the proof in Appendix B. 

\vglue0.3cm
\\There exist  positive definite  $C^{\infty}$  functions $\G_{c*} (\cdot, s)$  in ${\math
R}^{d}$ of finite range $L$ such that 

$$|| \dpr _{c} ^{p} \G_{c*} (\cdot, s) ||_{L^{\infty}
({\math R}^{d})} \le c_{L,p}   (1+s)^{-{2}}   \Eq(3.41) $$

\\where $\dpr _{c}^{p}$ are  continuum partial derivatives of order
$p$. Moreover we have the convergence estimate as $j \rightarrow \infty$

$$||\dpr _{\e_j} ^{p} \G_{j} (\cdot, s)
-\dpr_{c}^{p}\G_{c*} (\cdot, s)||_{L^{\infty}((\e_q{\math
 Z})^{d})}  \le  c_{L,p}   (1+s)^{-{2}}     L^{-{j\over2}}    \Eq(3.5) $$

\\and the continuum partial derivatives are taken in the same
 directions as the lattice partial derivatives.
This is Corollary 2.2 of [10] except that we have replaced the
exponential estimate in $\sqrt s$ by a power law estimate.

\vglue0.3cm 

\\{\it Remark 1}: In Appendix A of [9] interior regularity estimates (like those of
Nirenberg and Agmon in the continuum) were obtained for the solution
of a lattice Dirichlet problem for the
minus lattice laplacian plus a  mass squared  parameter (called $a\ge
0$). This is called $s$ in the present paper.
As part of this estimate a linear decay in the mass squared parameter  
was given and this sufficed for the purposes of [9]. However at the
end of Appendix A  [9] an exponential
type decay in the mass parameter was sketched following an Agmon type
argument. But on a lattice this will not be true for an
arbitrarily large mass parameter. However the exponential estimates are not necessary
as we will now see. We have replaced it by a weaker power law
decay which will suffice for our purpose.

\vglue0.3cm

\\{\it Remark 2}: The finite range expansion \equ(3.2) together with the 
bounds \equ(3.4), \equ(3.40), \equ(3.41) and \equ(3.5) 
remain valid in $d=2$ provided $s>0$. 

\vglue0.3cm

\\{\it Remark 3}:  Similar results are due to Bauerschmidt [5] by
 different methods. The constants in [5] are independent of $L$ except
 in $d=2$ where a $\log L$ dependence occurs. Moreover
the convergence rate is $L^{-j}$ instead of
 $L^{-{j\over 2}} $ as above.

\vglue0.3cm

\\{\it Remark 3}:
At the beginning of Theorem 1.1  we restricted $L$ to be
$L=3^{p}$. It was shown in [10] that the results of [9] which were
obtained under the condition $L=2^{p}$ continue to hold if $L=3^{p}$
which is useful for lattice RG applications. However such
 restrictions are unnecessary if one employs the methods of [5]. It
 would be sufficient to have the weaker condition $L\ge 2$.

\vglue0.3cm

\\Let

$$t= -\hat \D_{{\math Z}^{d}} (k)  \Eq(2.1) $$ 

\\where $\hat \D_{{\math Z}^{d}}$ is the Fourier transform of the
lattice Laplacian in ${\math Z}^{d}$  and $k\in [-\p, \p]^{d}$.
The Fourier transform of \equ(2.3) of Proposition 2.1 gives \equ(1.52)

$$ G_{\a}(x-y,m^{2})= \int_{0}^{\infty} ds\>   \r_{\a} (s, m^{2}) G(x-y,s)  $$

\\We insert on the right hand side the finite range decomposition of \equ(3.2) to get

$$G_{\a} (x-y, m^{2}) = \sum_{j\ge 0} L^{-j(d-2)} \int_{0}^{\infty} ds\>
\r_{\a}(s, m^{2}) \>  \G_{j} ({x-y\over L^{j}},\>
L^{2j} s ) $$ 

\\the interchange of sum and integral being permitted by virtue of the 
bounds  in \equ(3.4) above and in \equ(2.5) of Proposition 2.1.  After
change of variables (rescaling in s) we get

$$G_{\a} (x-y, m^{2}) = \sum_{j\ge 0} L^{-2j[\f]} \int_{0}^{\infty} ds\>
\r_{\a,j}(s, m^{2}) \>  \G_{j} ({x-y\over L^{j}},\>s ) \Eq(3.6)$$ 

\\where 

$$[\f]= {d-\a\over 2} $$

\\and

$$\r_{\a,j}(s, m^{2})= {1\over L^{j\a}} \> \r_{\a}({s\over L^{2j}}  ,
m^{2})$$

\\Explicit computation using the expression \equ(2.4) of Proposition 2.1 now gives

$$\r_{\a,j}(s, m^{2})= \r_{\a}(s, L^{j\a}m^{2}) \Eq(3.7)$$

\\Define the functions 

$$\G_{j,\a} (\cdot,\>m^{2} ) :\>(\e_{j} {\math Z} )^{d} \rightarrow {\math R}$$

\\by  

$$ \G_{j,\a} (\cdot,\>m^{2} )= \int_{0}^{\infty} ds\>\r_{\a}(s,\>  m^{2})\> \G_{j} (\cdot,\>s ) \Eq(3.8)$$

\\Note that  the functions $\G_{j,\a}(\cdot,\>m^{2} )$ are positive definite and of finite range $L$
because of the known properties of $\G_{j}$ stated after
\equ(3.3). They are well defined because of the bounds \equ(3.4) above
and \equ(2.5) of Proposition 2.1.

\\From \equ(3.6), \equ(3.7) and \equ(3.8) we get the desired finite
range decomposition

$$G_{\a} (x-y, m^{2}) = \sum_{j\ge 0} L^{-2j[\f]} \>\G_{j,\a}
({x-y\over L^{j}},\> L^{j\a} m^{2} ) \Eq(3.9)$$ 

\\stated in \equ(1.9) of Theorem 1.1. 

\vskip0.3cm

\\We will now prove the bounds \equ(1.10), \equ(1.101) and \equ(1.112) of Theorem 1.1.  

\vglue.03cm

\\From \equ(3.7), and the bounds \equ(3.4) and \equ(2.5) we get for
$j\ge 2$ with $0\le q\le j$

$$|| \dpr _{\e_j} ^{p} \G_{j,\a} (\cdot, m^{2}) ||_{L^{\infty}
((\e_q{\math Z})^{d})} \le  c_{L, p} c_{\a} \> \int_{0}^{\infty} ds \>
 {s^{\a/2} \over s^{\a} + m^{4}} \>
 (1+s)^{-2} \Eq(3.10) $$

\\Consider the integral on the right hand side of \equ(3.10) 

$$F(m^{2}) =\int_{0}^{\infty} ds \> {s^{\a/2} \over s^{\a} + m^{4}} \>
(1+s)^{-2}  \Eq(3.11)  $$

\\as a function of $m^{2}$. Note that the integral converges uniformly in
$m^{2}$ for $0<\a<2$. $F$ is a continuous monotonic increasing function of
$m^{2}$ as $m^{2}$ decreases and for $m^{2}=0$ we have

$$F(0) =\int_{0}^{\infty} ds \> s^{-\a/2} \>
 (1+s)^{-2}  = c_{1,\a}  \Eq(3.12)  $$

\\which is a constant of $O(1)$. For $m^{2}\neq 0$ we have 

$$F(m^{2}) \le m^{-4}  \int_{0}^{\infty} ds \> s^{\a/2} \>
(1+s)^{-2}  \le  c_{2,\a} \> 
m^{-4}  \Eq(3.13) $$

\\where $c_{2,\a} $ is a constant of $O(1)$. By continuity, there
exists a constant $c_{3, \a}$ independent of $m^{2}$ such that 

$$F(m^{2}) \le c_{3,\a}\> (1+ m^{4})^{-1}  \le 2c_{3,\a} \>    (1+m^{2})^{-2}  \Eq(3.14) $$

\\From \equ(3.10), \equ(3.11) and \equ(3.14) we get 

$$|| \dpr _{\e_j} ^{p} \G_{j,\a} (\cdot, m^{2}) ||_{L^{\infty}
((\e_q{\math Z})^{d})} \le  c_{L, p,\a} \> (1+m^{2})^{-2}    \Eq(3.15) $$

\\which proves \equ(1.10) of Theorem 1.1 when $j\ge 2$ with $0\le q\le
j$.

\vglue0.3cm
\\ For $j=0, 1$ we proceed otherwise to prove the bound \equ(1.101)

\vglue0.2cm

\\From \equ(3.40), \equ(3.7), \equ(3.8) and the bound \equ(2.5) we get
$$|| \dpr _{\e_j} ^{p} \G_{j,\a} (\cdot, m^{2}) ||_{L^{\infty}
((\e_q{\math Z})^{d})} \le  c_{L,p} c_{\a} \> \int_{0}^{\infty} ds \> {s^{\a/2} \over s^{\a} + m^{4}} \>
  {1\over 1+s}     \Eq(3.10) $$

\\Define the integral above as

$$F_{0}(m^{2}) =\int_{0}^{\infty} ds \> {s^{\a/2} \over s^{\a} + m^{4}} \>
{1\over 1+s}    \Eq(3.11)  $$

\\The integral converges  for $0<\a<2$. This is a continuous monotonic
increasing function for decreasing
 $m^2$ and is well defined for 
$m^2=0$ in the above range of $\a$. For $m^{2}\neq 0$ we obtain after
some changes of variables

$$\eqalign{ F_{0}(m^{2}) = & {1\over m^2}  (m^{2})^{2\over \a} {2\over \a} 
\int_{0}^\infty dx \> {e^{{2\over \a} x} \over e^x +e^{-x} } 
{1\over 1+ (m^{2})^{2\over \a} e^{{2\over \a} x} } \cr
&\le  {1\over m^2} {2\over \a} \int_{0}^\infty dx\> e^{-x} \cr
&\le   {2\over \a}  {1\over m^2} \cr}$$

\\By continuity at $m^{2} =0$ we have for some constant $c_{\a}$ 

$$F_{0}(m^{2}) \le c_{\a} {1\over 1+m^2}  \Eq(3.111)$$

\\Using this  bound in \equ(3.10) proves the bound 
\equ(1.101) for $j=0,1$.

\vglue0.3cm

\\The proof of \equ(1.11) and \equ(1.112) goes along similar lines.  The continuum
positive definite functions $\G_{c*,\a} (\cdot,\>m^{2} )$ of finite
range $L$ are defined by

$$ \G_{c*,\a} (\cdot,\>m^{2} )= \int_{0}^{\infty} ds\>\r_{\a}(s,\>  m^{2})\> \G_{c*} (\cdot,\>s ) \Eq(3.16)$$

\\Using the bound \equ(3.41) and the bound \equ(2.5) of Proposition
2.1 we get following the same chain of arguments as before the bound

$$|| \dpr _{c} ^{p} \G_{c*,\a} (\cdot, m^{2}) ||_{L^{\infty}
({\math R}^{d})} \le  c_{L, p,\a} \> (1+m^{2})^{-2} \Eq(3.17)$$ 

\\Now

$$\G_{j,\a} (\cdot,\>m^{2} )- \G_{c*,\a} (\cdot,
m^{2})=\int_{0}^{\infty} ds\>\r_{\a}(s,\>  m^{2})\> (\G_{j}
(\cdot,\>s) - \G_{c*} (\cdot,\>s )) \Eq(3.18) $$

\\whence we get  using the bounds \equ(2.5) and \equ(3.5)  and
following the earlier arguments for $j\ge 2$ sufficiently  large

$$||\dpr _{\e_j} ^{p} \G_{j,\a} (\cdot, m^{2})
-\dpr_{c}^{p}\G_{c*} (\cdot, m^{2})||_{L^{\infty}((\e_q{\math
 Z})^{d})}  \le  c_{L, p, \a} \>  (1+m^{2})^{-2} \>  L^{-{j\over2}}    \Eq(3.19) $$

\\The proof of Theorem 1.1 is complete. \bull

\vglue0.3cm

\\{\bf Appendix A: Coarse graining estimate}

\vglue0.3cm

\\In section 1, after the Corollary 1.2, coarse graining was
introduced, \equ(1.1411) and \equ(1.142). We claimed that the coarse
grained fluctuation covariances $\tilde \G'_{j,\a} (\cdot,m^{2})$
defined in \equ(1.1411) obey the same estimates as $\tilde \G_{j,\a} (\cdot,m^{2})$   with constants independent of
the coarse scale $L'=L^{r}$ for fixed initial scale $L$ for $d\ge 2$. We now
prove this:

\vglue0.3cm

\\{\it Proof}:  Let $e(j)=2:\> \forall \> j\ge 2$, and $e(j)=1 \>{\rm for}\>
j=0, 1$. Then from \equ(1.1411), \equ(1.14) and \equ(1.141) we get

$$ \eqalign{||\dpr_{{\math Z} ^{d}}^{p} \tilde \G'_{j,\a} (\cdot,
m^{2})||_{L^{\infty} ({\math Z}^{d})} &\le \sum_{l=0}^{r-1} ||\dpr_{{\math Z} ^{d}}^{p} \tilde \G'_{l+jr,\a} (\cdot,
m^{2})||_{L^{\infty} ({\math Z}^{d})} \cr
&\le c_{L, p, \a} \sum_{l=0}^{r-1} (1+ L^{(l+jr)\a} m^{2})^{-e(j)} L^{-(2[\f] +p)(l+jr)} \cr 
&\le c_{L, p, \a} (1+ (L^{r})^{j\a} m^{2})^{-e(j)} (L^{r})^{-(2[\f] +p)j}
\sum_{l=0}^{r-1} L^{-2l[\f]} \cr
&\le c_{L, p, \a}(1+ (L')^{j\a} m^{2})^{-e(j)} (L')^{-(2[\f] +p)j}
\sum_{l=0}^{\infty} L^{-2l[\f]}
\cr}$$

\\The last sum converges for $L\ge 2$ since $2[\f]= (d-\a) >0$ for
$d\ge 2$ and $0<\a<2$. 
Thus we obtain 

$$||\dpr_{{\math Z} ^{d}}^{p} \tilde \G'_{j,\a} (\cdot,
m^{2})||_{L^{\infty} ({\math Z}^{d})} \le c'_{L, p, \a}(1+ (L')^{j\a}
m^{2})^{-e(j)} (L')^{-(2[\f] +p)j} \Eq(A1)$$

\\where

$$c'_{L, p, \a}=c_{L, p, \a}(1- L^{-2[\f]}) \Eq(A2)$$ 

\\\equ(A1) thus gives the same bounds as \equ(1.10) and
\equ(1.101). Moreover \equ(A2) shows that the constant
is independent of $L'$.  This proves our claim. \bull
\vglue0.3cm

\\{\bf Appendix B: Proof of \equ(3.4)} 

\vglue0.3cm
\\In this appendix we will prove the bound in \equ(3.4), namely for
$j\ge 2$ 

$$|| \dpr _{\e_j} ^{p} \G_{j} (\cdot, s) ||_{L^{\infty}
((\e_q{\math Z})^{d})} \le   c_{L, p}   (1+s) ^{-2}  \Eq(B1)  $$

\\It can be proved in the same way as Theorem 5.5 of [9]. We will only indicate
the changes. Namely
take the Fourier transform of the formulae (3.28), (3.29) in [9] to get
after some change of notations (in particular $k$ is the Fourier variable)

$$\hat \Gamma_{j} (k,s) = |{\cal A}_{j} (k,s)|^{2} \hat \Gamma_{\e_{j}}(k,s) \Eq(B2)$$
\\where

$$\hat {\cal A}_{j} (k,s)=\prod_{m=1}^{j} \hat A_{\e_j, m}
(L^{-(m-1})(k,s) \Eq(2)$$ 

\\For $m=1, 2$ we have the bounds with $p'$ an arbitrary positive integer

$$|\hat A_{\e_j, m} (L^{-(m-1})(k,s)| \le
c_{L,p'} (1+s)^{-{1\over 2}} (k^{2} +1)^{-p'}, \> \forall\> p'\ge 0 \Eq(B3)$$

\\The case $m=1$  is covered in the proof of Lemma 5.4, [1]. The
case $m=2$ is proved in the same way following the same chain of arguments.
 As in the proof of Theorem 5.5, [1],  but now 
for $m\ge 3$ we use the bound  $| \hat A_{\e_j, m}(k,s)| \le
1$. Therefore for all $j\ge 2$ 

$$ |\hat{\cal A}_{j} (k,s)|^{2} \le c_{L,p'} (1+s)^{-{2}} (k^{2} +1)^{-p'}, \> \forall\> p'\ge 0 \Eq(B4)$$
 
\\Combining this with the uniform bound  
$|\hat\Gamma_{\e_{j}}(k,s)|\le c_{L}(1+k^{2})^{-1} $ 
given in the proof of Theorem 5.5 in[9] plus Sobolev embedding for
large enough $p'$ gives
us as (in the proof of Theorem 5.5)     \equ(3.4).  \bull

\vglue0.3cm

\\{\bf Acknowledgement}  The author is grateful to  Andr\'e Neveu for very
helpful conversations which led to the discovery and the original proof of Proposition
2.1 as well as J\'erome Dorignac and Jean-Charles Walter for supplying the figures. He
thanks David Brydges for numerous comments on an earlier version.  He
thanks Gordon Slade for bringing to the author's attention reference
[25] as well as a number of errors in [26] which were corrected in
[27] and has led to the present revised version.
He thanks two reviewers of the earlier version for
suggesting that the proof of Proposition 2.1 be shortened and making a
constructive suggestion to that end.

\vglue0.5cm

\\{\bf References}

\vglue0.3cm
\\[1] A. Abdesselam: A Complete Renermalization Group Trajectory
Between Two Fixed Points,
Commun. Math. Phys. (2007)  {\bf 276}  727-772

\vglue0.3cm
\\[2] : Aizenman and Fernandez: Critical Exponents 
for Long Range Interactions,

\\Lett.Math.Phys. (1988) {\bf 16} 39-49

\vglue0.3cm
\\[3] A. Abdesselam, A. Chandra and G. Guadagni: Rigorous Quantum
Field Theory Functional Integrals over the $p$-Adics $\bf 1$:
Anomalous Dimensions,  Preprint 2013

\\http://arxiv.org/pdf/1302.5971.pdf

\vglue0.3cm
\\[4] Stefan Adams, Roman Koteck\'y and Stefan M\"uller: Finite range range
decomposition for families of gradient Gaussian measures,
J. Funct. Anal. (2013) {\bf 264} 169-206 

\vglue0.3cm

\\[5] Roland Bauerschmidt: A simple method for finite range decomposition
of quadratic forms and Gaussian fields,
Probab. Theory Relat. Fields (2013) {\bf 157}: 817-845

\vglue0.3cm
\\[6] David Brydges: The Renormalisation Group and Self-Avoiding Walk:
In: Random Walks, Random Fields and Disordered Systems,  
Lecture Notes in Mathematics (2015) {\bf 2144} 65-116

\vglue0.3cm
\\[7] Roland Bauerschmidt, David C. Brydges and Gordon Slade:
Logarithmic Correction for the Susceptibility of the $4$-Dimensional
Weakly Self-Avoiding Walk: A Renormalisation Group Analysis,
Commun. Math. Phys. (2015) {\bf 337}: 817-877

\vglue0.3cm
\\[8] Roland Bauerschmidt, David C. Brydges and Gordon Slade:
Scaling Limits and Critical Behaviour of the $n$-component
$|\phi|^{4}$ Spin Model,
J. Stat. Phys. (2014) {\bf 157}: 692-742

\vglue.3truecm  
\\[9] D. Brydges, G. Guadagni and P. K. Mitter: Finite range
Decomposition of Gaussian Processes,
J. Stat. Phys. (2004) {\bf 115}: 415--449  

\vglue0.3cm

\\[10] David C Brydges and P. K. Mitter: On the Convergence to the
Continuum of Finite Range Lattice Covariances,
J. Stat. Phys. (2012) {\bf 147}: 716-727

\vglue0.3cm
\\[11] D. C. Brydges, P. K. Mitter and B. Scoppola: Critical
$\phi^{4}_{3,\e}$,
Commun.Math. Phys. (2003) {\bf 240}: 281-327 

\vglue.3truecm
\\[12] D. Brydges and A. Talarczyk: Finite Range Decomposition of
Positive Definite Functions,
J Funct Anal (2006)  {\bf 236}: 682-711 

\vglue0.3cm
\\[13] J. Dimock: Infinite Volume Limit for the Dipole Gas,
J. Stat. Phys. (2009) {\bf 135}: 393-427

\vglue0.3cm

\\[14] Pierluigi Falco: Kosterlitz-Thouless Transition Line for the Two
Dimensional Coulomb Gas,
Commun. Math. Phys. (2013) {\bf 312}: 559-609

\vglue0.3cm
\\[15] Pierluigi Falco: Critical Exponents of the Two Dimensional
Coulomb Gas at the Berezinki- Kosterlitz- Thouless Transition,
http://arxiv.org/pdf/1311.2237v2  (2013) 

\vglue0.3cm
\\[16] K. Gawedzki and A. Kupiainen: Massless ${\f}^{4}_{4} $:
Rigorous control of a renormalizable asymptotically free model,
Commun. Math. Phys. (1985) {\bf 99}: 197-252

\vglue0.3cm
\\[17] Takashi Hara: A Rigorous Control of Logarithmic Corrections in
Four-Dimensional $\phi^{4}$ systems. 1. Tragectory of Effective Hamiltonians,
J. Stat. Phys. (1987) {\bf 47}: 57- 98

\vglue0.3cm
\\[18] Takashi Hara and Hal Tasaki: A Rigorous Control of Logarithmic Corrections in
Four-Dimensional $\phi^{4}$ systems. II. Critical behavior of
Susceptibility and Correlation Length,
J. Stat. Phys. (1987) {\bf 47}: 99- 121

\vglue0.3cm
\\[19] P. K. Mitter: Long Range Ferromagnets: Renormalization Group
Analysis, 
\vglue0.1cm
https://hal.archives-ouvertes.fr/cel-01239463

\vglue0.3cm
\\[20] P. K. Mitter and B. Scoppola: The Global Renormalization Group
Trajectory in a Critical Supersymmetric Field Theory on the lattice
${\math Z}^{3}$,
J. Stat. Phys. (2008) {\bf 133}: 921- 1011

\vglue0.3cm
\\[21] K. G. Wilson and J. Kogut: Renormalization Group and the $\e$
expansion,
Phys. Rep. C (1974)  {\bf 12}: 75-200

\vglue0.3cm
\\[22] K. G. Wilson and J. Kogut: Feynman-Graph Expansion for Critical Exponents,
Phys. Rev. Lett. (1972) {\bf 28}: 548-550

\vglue0.3cm
\\[23] M. E. Fisher, S-K. Ma and B. G. Nickel: Critical Exponents for
Long-Range Interactions, 
Phys. Rev. Lett. (1972) {\bf 29}: 917-920

\vglue0.3cm

\\[24] E. T. Whittaker and G. N. Watson: A Course of Modern Analysis,
Cambridge University Press, Cambridge, England (1963): 244-245

\vglue0.3cm
\\[25] K. Yosida: Functional Analysis, Springer- Verlag, Berlin
Heidelberg New York (1978), Fifth Edition, page 260, formula 6.

\vglue0.3cm
\\[26] P. K. Mitter: On a finite range decomposition of the resolvent
of a fractional power of the laplacian,
J Stat Phys (2016) {\bf 163}:1235-1246

\vglue0.3cm
\\[27] P. K. Mitter: Errata:  On a finite range decomposition of the resolvent
of a fractional power of the laplacian,
J Stat Phys (to be published)

\vglue0.3cm
\\[28]  G. Slade: Critical exponents for long range $O(n)$ models below the
upper critical dimension, https://arxiv.org/1611.06169

\vglue0.3cm
\\[29] David C. Brydges: Lectures on the renormalisation group. In
{\it Statistical Mechanics}, volume 16, {IAS/Park City Math. Ser.,}
pages 7-93. Amer. Math. Soc., Providence, RI, 2009.

\vglue0.3cm
\\[30] R. Bauerschmidt (unpublished)

\bye